\documentstyle[aps,prb]{revtex}

\begin{document}
\title{Paramagnetic limit of superconductivity 
in a crystal without inversion center}
\author{V.P.Mineev}
\address{Commissariat a l'Energie Atomique, Departement de Recherche
Fondamentale sur la Matiere Condensee, SPSMS, 38054 Grenoble, France }
\date{ \today }
\maketitle
\begin{abstract}
The theory of paramagnetic limit of superconductivity in metals 
without inversion center is developed.  There is in general the
paramagnetic suppression of superconducting state.  The effect is
strongly dependent on field orientation in respect to crystal axes. 
The reason for this is that the degeneracy of electronic states with
opposite momenta ${\bf k}$ and $-{\bf k}$ forming of Cooper pairs is
lifted by magnetic fields but for some field directions this lifting
can be small or even absent.
\end{abstract}
 
\bigskip


\bigskip 
Quite recently the first unconventional superconductors without inversion
symmetry $CePt_{3}¥Si$ \cite{1} and $UIr$  \cite{2} have been discovered. 
The former reveals superconductivity in antiferromagnetic state \cite{3}
while the second is a ferromagnetic superconductor.  The microscopic
theory of superconductivity in metals without inversion has been
developed by V.Edel'stein \cite{4} pretty long ago.  The different
aspects of theory of superconductivity in such type materials has been
discussed about the same time \cite{5,6,7} and has been advanced
further in more recent publications \cite{8,9,10,11,12,13,14,15}. 
Finally, the general symmetry approch to the superconductivity in the
materials with space parity violation has been developed \cite{16,17}.

Particular attention has been attracted to the question about
paramagnetic limit in such type materials.
This problem has been treated by the authors \cite{11}, \cite{13} (in
2D metal with Rashba Hamiltonian) and quite recently in \cite{18}
(in 3D metal) by means of calculation of susceptibility in
superconducting state.  Another words it was done in the limit of
negligibly small magnetic field at finite value of the order
papameter.  Being useful for the establishing of the Knight shift the
susceptibility is not directly related to the paramagnetic limit
determination.  The latter has to be properly calculated in the limit of the
negligibly small order parameter at finite magnetic field.
That was undertaken in the paper \cite{15}.  

It was occured that zero
temperature upper critical field in polycrystilline $CePt_{3}¥Si$ is
about 5 Tesla \cite{1}, meanwhile the simple estimation of
paramagnetic limiting field $H_{p}¥=\pi T_{c}¥/\gamma\sqrt{2}\mu_{B}¥$
through the value of critical temperature $T_{c}¥=0.75K$ gives
$H_{p}¥\approx 1T$.  This observation is incompatible with
spin-singlet pairing and rather signals the spin-triplet
superconductivity.  The situation is even worse in $UIr$ where
superconductivity coexists with ferromagnetism.  The big internal
field in ferromagnetic metal moves apart the Fermi surfaces of the
bands filled by electrons with opposite spins making the singlet
pairing impossible.  On the other hand it is known \cite{4} that the
simple division on spin singlet and spin triplet pairing states does
not work in the crystals without inversion.

Hence, the problem of the paramagnetic limit in superconductors without
inversion deserves a special investigation and it was undertaken in
the paper \cite{15}.  From our point, this paper contains the inconsistency:
after the proper description of spinor electronic states in normal
metal without inversion , the authors introduce the superconducting pairing
interaction as in usual BCS theory for the crystals with inversion. 
So, they impose the pairing interaction between the states which do
not exist in normal state.  This point of view is may be acquited in
the crystal with negligibly small spin-orbital coupling having no
influence on the pairing interaction as it has been considered in the
original paper \cite{4}.  However, in general, the assumption, that
pairing takes place between the states which are not modified by the
absence of the inversion center, is equivalent to the assumption that
typical for the metal without inversion and odd on electronic momentum
spin-orbital coupling is smaller than superconducting critical
temperature $T_{c}¥$.  This point of inconsistency is absent in the
papers \cite{16,17} where the general symmetry approach to the problem
of supperconductivity in the crystal without inversion has been
developed.  There was shown in particular \cite{16} that the band
splitting due to the lack of inversion in $CePt_{3}¥Si$ cannot at all be
considered as small.  Hence from our point of view the problem of
paramagnetic limit raised in \cite{15} must be reconsidered and we do
it in the present article.

It is shown that the paramagnetic suppression of superconducting state
in a crystal without inversion centrum certainly exists and the effect
is strongly dependent of field orientation in respect of crystall
axes.  Whereas in general the paramagnetic limiting
field is roughly the same as in a singlet superconductor, for some
field directions $H_{p}¥$ is very large or even infinite.  These
are those directions where the magnetic field lifting of degeneracy of
electronic states with opposite momenta ${\bf k}$ and $-{\bf k}$
forming the Cooper pairs is absent.

Let us start from description of normal state in the crystal without
inversion centrum.  For each band its single-electron Hamiltonian has
the form
\begin{equation}
H=\varepsilon^{0}¥_{{\bf k}}¥+ \bbox{\alpha}_{{\bf k}}¥\bbox{\sigma},
\label{e1}
\end{equation}
where ${\bf k}$ is the wave-vector, the 
$\varepsilon^{0}¥_{{\bf k}}¥=\varepsilon^{0}¥_{-{\bf k}}¥$ is even function
of ${\bf k}$, $\bbox{\alpha}_{{\bf k}}¥= -\bbox{\alpha}_{-{\bf k}}¥$
is odd pseudovectorial function of ${\bf k}$,
$\bbox{\sigma}=(\sigma_{x}¥,\sigma_{y}¥,\sigma_{z}¥)$ is the vector
consisting of Pauli matrices.  The eigen values and eigen functions of
this Hamiltonian are
\begin{equation}
\varepsilon_{{\bf k}\lambda }¥=\varepsilon^{0}¥_{{\bf k}}¥-\lambda
|\bbox{\alpha}_{{\bf k}}¥|,
\label{e2}
\end{equation}
\begin{eqnarray}
\Psi_{\lambda }¥({\bf k})\propto \left(
\begin{array}{c}
-\alpha_{{\bf k}x}¥+i\alpha_{{\bf k}y}¥ \\
\alpha_{{\bf k}z}¥+\lambda|\bbox{\alpha}_{{\bf k}}¥| \end{array}\right).
\label{e3}
\end{eqnarray}
So, we have obtained the band splitting and $\lambda=\pm $ is the band index.
As result, there are two Fermi surfaces determined by equations
\begin{equation}
\varepsilon_{{\bf k}\lambda }¥=\varepsilon_{F}¥,
\label{e4}
\end{equation}
which may of course have the degeneracy points or lines for some
directions of ${\bf k}$.  The symmetry of directions of the dispersion laws
$\varepsilon_{{\bf k}\lambda }¥$ has to correspond to the crystal symmetry.
Particular attention however deserves the operation of reflection ${\bf k}$ to
$-{\bf k}$ which creates the time reversed states.

By application of operator of time
inversion $\hat K=-i\sigma_{y}¥K_{0}¥$, where $K_{0}¥$ is the
complex-conjugation operator one can see that the state $\Psi_{\lambda
}¥({\bf k})$ and the state inversed in time $\hat K \Psi_{\lambda
}¥({\bf k})\propto\Psi_{\lambda}¥(-{\bf k})$ are degenerate.  Another
words, they correspond to the same energy $\varepsilon_{{\bf k}\lambda
}¥=\varepsilon_{-{\bf k}\lambda }¥$.  So, the Fermi
surfaces in a crystal without inversion center still have mirror symmetry.
This is the consequence of time inversion symmetry.

Let us look now on the modifications which are appeared by the application of
external magnetic field.  It is known \cite{19} that the field introduction
in Hamiltonian is made by the Peierls substitution ${\bf k}\to{\bf
k}+(e/2\hbar c){\bf H}\times(\partial/\partial{\bf k})$.  Being
interested in paramagnetic influence on superconductivity and
considering only the fields values $\mu_{B}¥H\ll\varepsilon_{F}¥$ one
can neclect by the term with magnetic field in the Peierls substitution
and take into account only direct paramagnetic influence of magnetic
field
\begin{equation}
H=\varepsilon^{0}¥_{{\bf k}}¥+ \bbox{\alpha}_{{\bf k}}¥\bbox{\sigma}
-\bbox{\mu}_{{\bf k}i}¥H_{i}¥\bbox{\sigma},
\label{e5}
\end{equation}
where $\bbox{\mu}_{{\bf k}i}¥=\bbox{\mu}_{-{\bf k}i}¥$ 
is even tensorial function of ${\bf k}$.  In the isotropic approximation
$\mu_{ij}¥=\mu_{B}¥g\delta_{ij}¥/2$, where $g$ is gyromagnetic ratio.
The eigen values of this Hamiltonian 
are
\begin{equation}
\varepsilon_{{\bf k}\lambda }¥=\varepsilon^{0}¥_{{\bf k}}¥-\lambda
|\bbox{\alpha}_{{\bf k}}¥-\bbox{\mu}_{{\bf k}i}¥H_{i}¥|.
\label{e6}
\end{equation}
It is obvious from here
that the time reversal symmetry is lost
$\varepsilon_{-{\bf k}\lambda }¥\ne\varepsilon_{{\bf k}\lambda }¥$ 
and the shape of the Fermi 
surfaces
do not obey the mirror symmetry.

If we have the normal one-electron states classification in a crystal
without inversion symmetry it is quite natural to describe the
superconductivity directly in the basis of these states.  So, the BCS
Hamiltonian in the space homogeneous case, which we discuss, looks as
follows
\begin{equation}
H_{BCS}¥=\sum_{{\bf k},\lambda}\xi_{{\bf k}\lambda }¥ a^{\dag}_{ {\bf
k}\lambda} a_{{\bf k}\lambda} +\frac{1}{2} \sum_{{\bf k},{\bf
k}',\lambda,\nu} V_{\lambda \nu}({\bf k},{\bf k}') a^{\dag}_{ -{\bf
k}, \lambda} a^{\dag}_{ {\bf k},\lambda} a_{ {\bf k}', \nu} a_{ -{\bf
k}', \nu},
\label{e7}
\end{equation}
where 
$\lambda, \nu=\pm$ are the band indices for the bands intoduced above
and 
\begin{equation}
\xi_{{\bf k}\lambda }¥=\varepsilon_{{\bf k}\lambda }¥-\mu
\label{e8}
\end{equation}
are the
band energies counted from the chemical potential.  Due to big
difference between the Fermi momenta we neglect in Hamiltonian by the
pairing of electronic states from different bands.  The structure of
theory is now very similar to the theory of ferromagnetic
superconductors with triplet pairing \cite{20}.  For Gor'kov equations
in each band we have
\begin{eqnarray}
\left(i\omega_{n}-\xi_{{\bf k}\lambda}\right) G_{\lambda}({\bf
k},\omega_{n})+ \Delta_{{\bf k}\lambda} F_{\lambda}^{\dagger}({\bf
k},\omega_{n})=1  \\
\left(i\omega_{n}+\xi_{-{\bf k}\lambda}\right) F_{\lambda}^{\dagger}¥ ({\bf
k},\omega_{n})+\Delta_{{\bf k}\lambda}^{\dagger}¥  G_{\lambda}({\bf
k},\omega_{n})=0,
\label{e9}
\end{eqnarray}
where $\omega_{n}¥=\pi T(2n+1)$ are Matsubara frequencies.
The equations for each band are only coupled through the order
parameters given by the self-consistency equations
\begin{equation}
\Delta_{{\bf k}\lambda}=-T\sum_{n}\sum_{{\bf
k}'}\sum_{\nu} V_{\lambda\nu}\left( {\bf k},{\bf k}'\right)
F_{\nu}({\bf k}',\omega_{n}).
\label{e10}
\end{equation}
The superconductor Green's functions are
\begin{eqnarray}
G_{\lambda}\left({\bf k},\omega_{n}\right) &=&
\frac{i\omega_{n}+\xi_{-{\bf k}\lambda}}
{(i\omega_{n}-\xi_{{\bf k}\lambda})(i\omega_{n}+\xi_{-{\bf k}\lambda})
-\Delta_{{\bf k}\lambda}\Delta_{{\bf k}\lambda}^{\dagger}¥ } \\
F_{\lambda}\left({\bf k},\omega_{n}\right)&=& \frac{-\Delta_{{\bf k}\lambda} 
}{(i\omega_{n}-\xi_{{\bf k}\lambda})(i\omega_{n}+\xi_{-{\bf k}\lambda})
-\Delta_{{\bf k}\lambda}\Delta_{{\bf k}\lambda}^{\dagger}¥ }.
\label{e11}
\end{eqnarray}
The energies of elementary excitations are given by
\begin{equation}
E_{{\bf k}\lambda}¥=
\frac{\xi_{{\bf k}\lambda}¥-\xi_{-{\bf k}\lambda}¥}{2}\pm\sqrt
{\left(\frac{\xi_{{\bf k}\lambda}¥+\xi_{-{\bf k}\lambda}¥}{2}
\right)^{2}¥+
\Delta_{{\bf k}\lambda}\Delta_{{\bf k}\lambda}^{\dagger}¥ }.
\label{12}
\end{equation}
 
For simplicity let us assume that we have pairing only in one band: 
$\lambda=+$.
The treatment of general case is similar but more lengthly.
There was shown in \cite{17} that in the case of crystals
without inversion:(i) $\Delta_{{\bf k}}¥=t({\bf
k})\sum_{i}¥\eta_{i}¥\varphi_{i}¥({\bf k})$, where $t({\bf k})=-t(-{\bf k})$
is an odd phase factor, 
(ii) a potential of the pairing interaction is represented
as an expansion over $t({\bf k})\varphi_{i}¥(\hat{\bf k})$, where
$\varphi_{i}¥(\hat{\bf k})$ are the even basis functions of an
irreducible representation of the crystal point symmetry group.  For
tetragonal crystal $CePt_{3}¥Si$ this group is $C_{4v}¥$ and for
monoclinic crystal $UIr$ it is $C_{2}¥$.  If we limited ourselves by
consideration only one-dimensional representations when we have
$V_{++}({\bf k},{\bf k}')=V~t({\bf k})t^{*}¥({\bf k}')\varphi(\hat{\bf
k})\varphi^{*}¥(\hat{\bf k}')$.

The equation for critical temperature that is the linear version of
(\ref{e10}) has in this case the form
\begin{equation}
1=-VT\sum_{n}\sum_{{\bf k}'}
\frac{\varphi^{*}¥(\hat{\bf k}') \varphi({\bf k}')}
{(i\omega_{n}¥-\xi_{{\bf k}'}¥)(-i\omega_{n}¥-\xi_{-{\bf k}'}¥)}.
\label{e13}
\end{equation}

Is clear from here and equations (\ref{e6}), (\ref{e8}) that the coherence
between the normal metal states with states with Green functions
$G^{0}¥({\bf k},\omega_{n})$ and $G^{0}¥(-{\bf k},-\omega_{n})$ is broken
by magnetic field.  The oppositely directed momenta ${\bf k}$ and
$-{\bf k}$ on the Fermi surface have the different length.  Hence the
magnetic field will suppress superconductivity that means the critical
temperature will be decreasing function of magnetic field.
It is clear also that it will be anisotropc function of the field
orientation in respect of cristallographic directions.  

For tetragonal crystal $CePt_{3}¥Si$ one can take as the simplest form
of gyromagnetic tensor $\mu_{ij}¥=\mu_{B}¥(g_{\perp}¥
(\hat x_{i}¥\hat x_{j}¥+\hat y_{i}¥ \hat y_{j}¥)+
g_{\parallel}¥\hat z_{i}¥\hat z_{j})/2$ and the pseudovector function
$\bbox{\alpha}_{{\bf k}}¥=\alpha(\hat z\times {\bf k})+ \beta\hat z
k_{x}¥k_{y}¥k_{z}¥(k_{x}¥^{2}¥-k_{y}¥^{2}¥)$. 
The latter is chosen following the discussion in the paper \cite{18}. 
Then for the normal metal energy of excitations we have
\begin{equation}
\xi_{\bf k}¥=\xi^{0}¥_{\bf k}¥-
\sqrt{(\alpha k_{y}¥+\frac{g_{\perp}¥}{2}\mu_{B}¥H_{x}¥)^{2}¥+
(\alpha k_{x}¥-\frac{g_{\perp}¥}{2}\mu_{B}¥H_{y}¥)^{2}¥+ 
(\beta k_{x}¥k_{y}¥k_{z}¥(k_{x}¥^{2}¥-k_{y}¥^{2}¥)
-\frac{g_{\parallel}¥}{2}\mu_{B}¥H_{z}¥)^{2}¥}
\label{e14}
\end{equation}

As result of simple calculation near $T_{c}¥$ we obtain
\begin{equation}
T_{c}¥({\bf H})=T_{c}¥\left\{1-
\frac{7\zeta(3)\mu_{B}¥^{2}¥}{32\pi^{2}¥T_{c}¥^{2}¥}
\left (a g_{\perp}¥^{2}¥(H_{x}¥^{2}¥+H_{y}¥^{2}¥)+
b g_{\parallel}¥^{2}¥H_{z}¥^{2}¥\right )+\ldots \right \},
\label{e15}
\end{equation}
that looks like similar to usual superconductivity with singlet pairing.
Here $a$ and $b$ are coefficients of the order of unity.  Its exact
values depend on the particular form of $\varphi(\hat{\bf k})$ functions
in pairing interaction as well on particular form of $\bbox{\alpha}_{{\bf k}}¥$.

On the other hand, let as assume that due to some particular reason 
coefficient $\beta$ is small.  Then for the field direction 
${\bf H}=H\hat z$ for $\mu_{B}¥g_{\parallel}¥H\gg\beta k_{F}¥^{5}¥$ we
have for the excitations energy
\begin{equation}
\xi_{\bf k}¥=\xi^{0}¥_{\bf k}¥-
\sqrt{(\alpha k_{y}¥)^{2}¥+ (\alpha
k_{x}¥)^{2}¥+ (\frac{g_{\parallel}¥}{2}\mu_{B}¥H_{z}¥)^{2}¥},
\label{e16}
\end{equation}
that is now the even function of the wave vector
$\xi_{\bf k}¥=\xi_{-{\bf k}}¥$ and the equation for the critical
temperature is
\begin{equation}
1= -VT\sum_{n}\int d\xi N_{\xi=0}¥(\hat {\bf k}') \frac{dS_{\hat {\bf
k}'}¥}{S_{F}¥}\frac{\varphi^{*}¥(\hat{\bf k}') \varphi({\bf k}')}
{(i\omega_{n}¥-\xi) (-i\omega_{n}¥-\xi)}.
\label{e17}
\end{equation}
Here we can first integrate over the energy variable
$\xi$ and and then over the Fermi suface.  After the first integration
the magnetic field dependence is disappeared from equation and we
obtain standart BCS formula $T_{c}¥=(2\gamma/\pi)\epsilon\exp(-1/g)$
for critical temperature determination.  So, the suppression of critical
temperature by magnetic field is saturated at finite value which
differs from its value at $H=0$ due to field variation of density of
states and pairing interaction at $\xi=0$.

This results can be in principle valid for any direction of magnetic field
if paramagnetic interaction exceeds a spin-orbital splitting
$|\bbox{\mu}_{i}¥H_{i}¥|>|\bbox{\alpha}|$. 
Of course the superconductivity in the region of the
large fields still exists if $g$ is positive on the Fermi surface
$\xi=0$.  Thus at large fields the situation is similar to that we
have in the supercoductors with triplet pairing.

We have demonstrated that the paramagnetic suppression of superconducting
state in a crystal without inversion centrum certainly exists and the
effect depends of field orientation in respect of crystall axes.  The
paramagnetic suppression of superconductivity takes place due to
magnetic field lifting of degeneracy of electronic states with
opposite momenta ${\bf k}$ and $-{\bf k}$ forming the Cooper pairs. 
For some directions of fields the degeneracy is recreated.  That is
why the paramagnetic limit of superconductivity in the crystals
without inversion can be in principle absent.

The similar conclusions have been obtained in the paper \cite{15}
in the assumption of negligibly small band splitting.  So, our main result 
is the development of proper theoretical treatment of  the paramagnetic
limitations of superconductivity in noncentrosymmetric metals with large band
splitting.

I am indebted to K.Samokhin who pointed out me on incorrect choice
of pseudovector $\bbox{\alpha}_{{\bf k}}¥$ in the first version of the article.

\end{document}